\documentclass[twocolumn,floatfix,aps,rmp]{revtex4}
\usepackage{graphicx}
\usepackage{amsmath}
\usepackage{amssymb}
\usepackage{bm}
\usepackage{siunitx}
\usepackage{setspace}
\usepackage[rightcaption]{sidecap}
\usepackage{url}
\usepackage[unicode,colorlinks=true,allcolors=blue]{hyperref}
\setcitestyle{numbers,comma,square,compress}
\begin{document}
\title{\large Room-temperature superconductivity in an artificial 2D Mott-insulating square lattice\\and its advanced condensed phase that generates a low-loss current in the atmosphere\\: A possible perpetual motion machine on the back of spontaneous symmetry breaking}
\author{Nobuyuki Zen}
\affiliation{Device Technology Research Institute, National Institute of Advanced Industrial Science and Technology,\\Tsukuba 305-8568, Ibaraki, Japan}
\begin{abstract}
A 2D metallic phononic crystal (PnC), which is fabricated by drilling periodic holes in a suspended niobium (Nb) film, is repeatedly cooled and warmed in the temperature range of 2--300 K. During the first five temperature cycles, the resistance of the metallic PnC gradually increases in accordance with the Friedel sum rule, indicating that narrow Nb bridges between adjacent thru-holes are converted into \emph{d}-orbital-filled Mott insulators. The consequent 2D Mott-insulating square lattice is a crystallographic analog of a copper oxide layer in high-temperature superconductors such as YBCO and BSCCO. Subsequent temperature cycles of the thus produced ideal Hubbard crystal realize zero resistance at 60 K, and the zero-resistance state remains up to 300 K, being retained in the atmosphere (i.e., at room temperature, in a terrestrial magnetic field, and under atmospheric pressure). The necessity of the latter temperature cycles remains unclear in this study; however, a possible transition mechanism involving the combined Josephson and charging effects is discussed. The critical current and critical field of the room-temperature superconductor (RTSC) are $I_{C}$ = 18.8 mA and $\mu_{0}H_{\perp}$ = 12 T, respectively. Once a large current exceeding $I_{C}$ is applied to the RTSC, some of the \emph{d} electrons are forced out, and a special interface consisting of the p-type semiconducting and superconducting states is formed. Carriers diffuse because of the nonequilibrium charge concentration, but they are Andreev-reflected back. In the reversible, thermally isolated system, the entropy never increases, and a simple experiment confirms a low-loss current generated from the interface spontaneously. That is, this study challenges the forbidden perpetual motion machine which can be a solution to the World energy crisis.
\medskip\\
{\tt keywords: metallic phononic crystal, room-temperature superconductivity,\\ambient pressure, semiconducting-superconducting interface, Andreev reflection,\\Majorana fermions, time's arrow, entropy, perpetual motion machine}
\end{abstract}

\maketitle

\small
\begin{spacing}{1.15}
\section{INTRODUCTION}
\label{sec:introduction}
\vspace{-0.2cm}
A total of 106 years has passed since Onnes discovered metal--superconductor transition at very low temperatures~\cite{Onn1911}. The BCS theory released in 1957, approximately half a century after Onnes’s discovery, revealed that the transition mechanism is a phonon-mediated electron attraction~\cite{Bar1957}. However, the discovery of a high-temperature superconductor (HTSC) by Bednorz and Müller in 1986~\cite{Bed1986} had profound repercussions for BCS theory. Specifically, rather than the high transition temperature ($T_{C}$), the problem was with the HTSC itself because it contains strong electron--electron (\emph{e}--\emph{e}) repulsion systems. Although the paradox of the attractive and repulsive forces might still be worth debating~\cite{And1984,And1987,And2007}, it has been established that \emph{d}-wave Cooper pairs are essential in HTSCs~\cite{Aok2014,Shi2016,Fuk2009,Sca1995} and condensed-matter physicists appear to have reached a consensus that HTSC can be achieved near the metallic and antiferromagnetic insulating phases. Very recently, room-temperature superconductivity is reported in a carbonaceous sulfur hydride under the extremely high-pressure limit~\cite{Sni2020}. However, such hydrides~\cite{Sni2020,Dro2015,Som2019,Sem2020} are BCS superconductors, the pairing mechanism of which is not correlated with that of superconductivity presented in this study.

According to the Hubbard model, an ideal crystal structure of the \emph{e}--\emph{e} repulsion system to achieve higher $T_{C}$ is a two-dimensional (2D) square lattice in which conductive and insulating regions are alternately arranged~\cite{Aok2014,Mon1999,Ari1999}. In fact, such an ideal Hubbard crystal (HbC) is inherently included in HTSCs as a copper oxide (CuO\textsubscript{2}) layer; a moderate balance between the repulsive interaction \emph{U} and the transfer integral \emph{t} in the CuO\textsubscript{2} plane realizes HTSC. Another material which might naturally include HbC is graphite, although its inhomogeneous nature, i.e., the spatially limited supercurrent path, results in a low superconducting yield~\cite{Sch2012,Pre2016}. The so-called granular superconductivity is well described by the nearest-neighbor charging model~\cite{Sim1994,Bra1984} which is analogous to the Hubbard model. To develop the principle, Tinkham considered only the \emph{2D square} array of Josephson junctions (JJs) in his textbook [see the Section 6.6 in Ref.~\cite{Tin2004} and Ref.~\cite{Phi1993}], which indicates the importance of the lattice structure itself and/or the fundamental structural element: the square closed circuit formed by four JJs.

\begin{SCfigure*}[0.4][t]
\includegraphics[width=12cm]{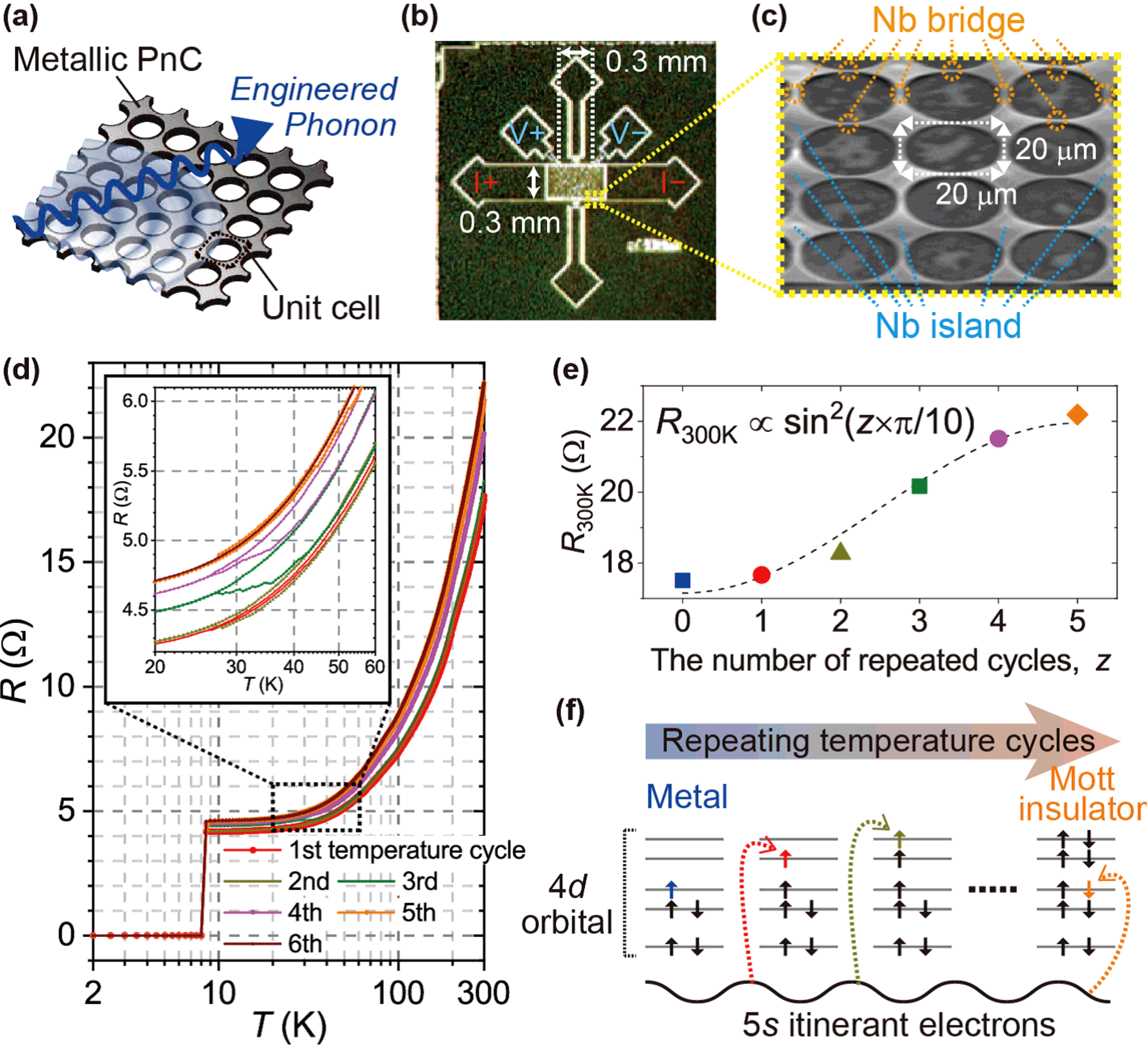}
\caption{\label{fig1}
(\textbf{a}) Schematic of a traveling phonon in a 2D phononic crystal (PnC). In case of metallic PnCs, the engineered phonons are expected to mediate abnormal 2D \emph{e}--\emph{e} interactions. (\textbf{b}) Optical micrograph of PnC--Nb used in this study; the periodically perforated area is PnC ($0.3\times0.5$ mm\textsuperscript{2}); the thickness is 150 nm, self-standing. The pathway of the excitation current (i.e., $I+$ toward $I-$) for \emph{R}--\emph{T} measurements is unidirectional, and the distance between electrical pads for voltage readings is 0.3 mm. (\textbf{c}) SEM of the region surrounded by the yellow dotted square in (b); the array of Nb islands and Nb bridges forms a 2D square lattice with a lattice constant of 20 $\si{\micro}$m. (\textbf{d}) \emph{R}--\emph{T} curves; only the initial six temperature cycles are shown here. Inset, zoomed-in plot. (\textbf{e}) \emph{R} at 300 K in (d) versus the number of temperature cycles, \emph{z}. (\textbf{f}) 4\emph{d}-orbital filled model, based on the Friedel sum rule indicated in (e); after the first five cycles, the Nb bridges undergo a transition to Mott insulators.
}
\end{SCfigure*}

Recently, we have shown that a periodically perforated niobium (Nb) film undergoes a transition into a Mott insulator~\cite{Zen2019}. While temperature cycling, 2D antiferromagnetic \emph{e}--\emph{e} interactions have been confirmed, and it is concluded that the Mott metal--insulator transition takes place at the narrow Nb bridges that connect adjacent conductive Nb islands. In general, such a periodically perforated structure is applied to a dielectric film in order to control its own thermal conductance~\cite{Maa2014,Mal2015}. The so-called phononic crystal (PnC) is schematically shown in Fig.~\ref{fig1}(a). In the metallic PnC, on the other hand, the forcibly created 2D Brillouin zone (BZ) of the phonon system affects the dimensionality of electron momenta and produces abnormal 2D \emph{e}--\emph{e} interactions~\cite{Zen2019}.

The study of metallic PnC systems has just begun. Nevertheless, the confirmed temporal stability of the fabricated 2D Mott-insulating lattice in the atmosphere in the previous study supports further development. Thus, this study begins by creating a 2D Mott-insulating square lattice: an ideal HbC that is the precursor of a room-temperature superconductor (RTSC). Unfortunately, investigating the properties of the RTSC was not straightforward. However, we continued investigation into its advanced condensed phase.

\section{RESULTS}
\label{sec:results}

\subsection{Mott transition of PnC--Nb to form HbC}
\label{sec:motttransition}
Figure~\ref{fig1}(b) shows a bird’s eye view of the PnC--Nb used in this study. The processed area is the PnC (i.e., a periodically perforated area of $0.3\times0.5$ mm\textsuperscript{2}), and the distance between the electrical pads for voltage readings in resistance \emph{vs.} temperature (\emph{R}--\emph{T}) measurements is 0.3 mm. Contrary to the previously reported Corbino disk geometry PnC--Nb sample~\cite{Zen2019}, the pathway of the excitation current for \emph{R}--\emph{T} measurements, which is denoted as $I+$ toward $I-$, is unidirectional. A scanning electron micrograph (SEM) of the region surrounded by the yellow dotted square of Fig.~\ref{fig1}(b) is shown in Fig.~\ref{fig1}(c). The periodic perforation creates a 2D square lattice of Nb islands with a lattice constant \emph{a} of 20 $\si{\micro}$m, and each Nb island is continuously connected by a narrow Nb bridge with a width of 200--300 nm. The thickness of the Nb film is 150 nm, and the film is suspended to prevent PnC phonons from being affected by the underlying layer [see ``Materials and Methods'' in the \hyperref[supplmater]{supplementary material} for details]. Regardless of the fully 3D film thickness for conducting electrons, the 2D BZ of the PnC results in the 2D phonon-mediated \emph{e}--\emph{e} interaction, as reported in the previous study~\cite{Zen2019}.

Figure~\ref{fig1}(d) shows the \emph{R}--\emph{T} results of the PnC--Nb sample; only the initial six temperature cycles are shown here. To create HbC, it is necessary to repeatedly cool and warm the PnC. Each temperature cycle consists of a cooling process from 300 K to 2K and a subsequent warming process from 2 K to 300 K. The resistance was measured by the four-probe method using electrical pads as denoted in Fig.~\ref{fig1}(b), and the peak amplitude of the applied excitation current was 10 $\si{\micro}$A. The \emph{R}--\emph{T} measurements were performed using the physical property measurement system (PPMS, Quantum Design) in zero magnetic field, and the sample space was kept at a low pressure of approximately 200 Pa. Before the beginning of the first cooling process, PnC--Nb was in the usual metallic state, and its resistance was 17.5 $\si{\ohm}$ at 300 K. In the first cooling process, the PnC--Nb undergoes an SC transition at 8.5 K, which is the usual $T_{C}$ for Nb, and it returns to the normal state at the same $T_{C}$ in the subsequent warming process. This feature is common to all temperature cycles. However, unlike other superconducting materials, the warming curve does not retrace the cooling curve during the first five cycles. The untraced feature is clearly seen in the third temperature cycle (green) in the inset of Fig.~\ref{fig1}(d) which shows the temperature range 20--60 K in detail. However, the untraced feature is not observed during the sixth cycle in which its warming curve retraces the cooling one [details can be referred from Fig. S1(b) in the \hyperref[supplmater]{supplementary material}].

The reason that the abnormal untraced feature only occurs in the first five cycles can be determined by plotting the increase in the resistance as a function of the number of temperature cycles \emph{z}, as shown in Fig.~\ref{fig1}(e). For clarity, the resistance at 300 K, $R_{300K}$ is used in plotting. $R_{300K}$ of \emph{z} = 0 is the resistance before the first cooling (17.5 $\si{\ohm}$). $R_{300K}$ is proportional to $\sin^{2}(z\times\pi/10)$, which indicates that PnC--Nb obeys the Friedel sum rule during these temperature cycles. In general, the Friedel sum rule is observed when a transient metal element is mixed as an impurity with a ground metal. The difference in the valence number \emph{z} between the ground metal and the mixture results in an increase in the resistance, as expressed in the abovementioned formula~\cite{Dat1997,Vil2016}. The number of the \emph{d} orbital is five, and two electrons with opposite spins can occupy each orbital. Figure~\ref{fig1}(e) indicates that some Nb atoms that are distributed in the PnC--Nb gain an additional electron during each temperature cycle; in other words, these atoms continue increasing in valence number, becoming different and abnormal Nb atoms. As schematically shown in Fig.~\ref{fig1}(f), after five temperature cycles, the abnormal Nb atoms undergo a transition to the Mott-insulating state in which the 4\emph{d} orbital is filled. The experimental tunneling current in Ref.~\cite{Zen2019} indicates that such insulating Nb atoms exist at the narrow Nb bridges that connect adjacent metallic Nb islands [see Fig.~\ref{fig1}(c)].

Thus, the 2D Mott-insulating square lattice, i.e., the ideal HbC, is created by repeatedly cooling and warming the PnC--Nb. However, it has a distinct difference from a native HbC (e.g., a CuO\textsubscript{2} plane in an HTSC); that is, the lattice constant of the artificial HbC is extraordinarily long, 20 $\si{\micro}$m.

\subsection{Subsequent temperature cycles of HbC to obtain RTSC}
\label{sec:rtsc}
\begin{figure}[h]
\includegraphics[width=7.5cm]{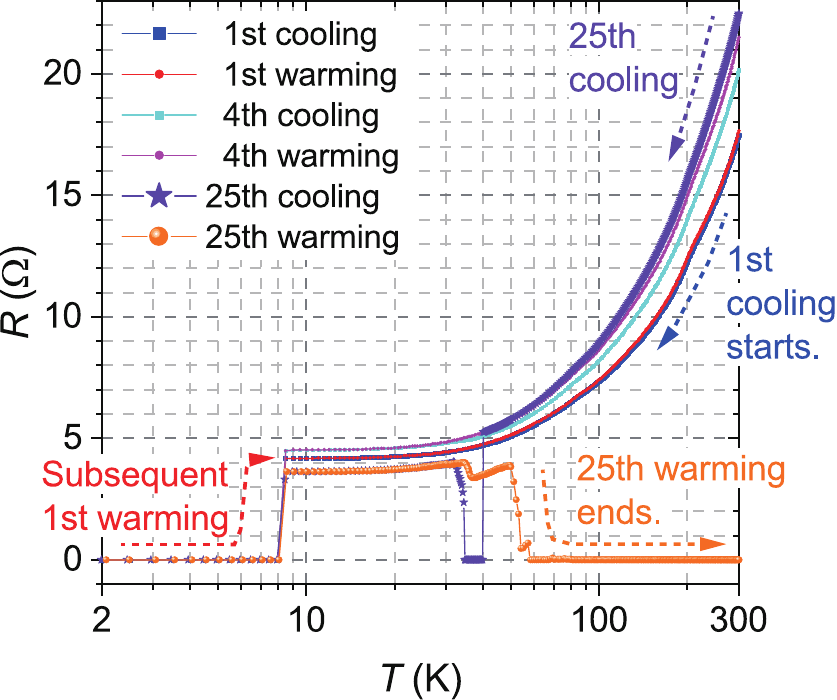}
\caption{\label{fig2}
\emph{R}--\emph{T} curves; only the first, fourth, and 25th cycles are shown. For all other \emph{R}--\emph{T} curves, see Fig. S1 in the \hyperref[supplmater]{supplementary material}.
}
\end{figure}

Subsequent temperature cycles of the thus created HbC realize zero resistance at 300 K. Figure~\ref{fig2} shows the \emph{R}--\emph{T} curve of the 25th temperature cycle; for comparison, the first and fourth curves, which were shown in Fig.~\ref{fig1}(d), are also included. All other \emph{R}--\emph{T} curves and their cooling--warming procedures are summarized in Fig. S1 and Table S1, respectively, in the \hyperref[supplmater]{supplementary material}; additionally, the experimental raw data is openly available in Dryad~\cite{Zen2020}. During the 25th cooling process (purple), the resistance suddenly disappears at 40 K, and after a while, a finite resistance appears again, undergoing a transition to an SC at 8.5 K. Finally, during the 25th warming process (orange), the resistance starts to decrease at 50 K and drops to zero at 60 K. This zero-resistance state remains up to 300 K. Unpredictably, RTSC originates from a low temperature.

The necessity for the subsequent temperature cycles after the sixth one will be discussed in the \hyperref[sec:iiib]{Section III. B} in connection with the \emph{combined Josephson and charging effects}~\cite{Ave1985,Ful1989} which must stem from competition between \emph{U} and \emph{t} in the capacitive Mott-insulating lattice.

\vspace{-1.0cm}

\subsection{Critical magnetic field of RTSC}
\label{sec:crimag}
\begin{figure}[h]
\includegraphics[width=6.8cm]{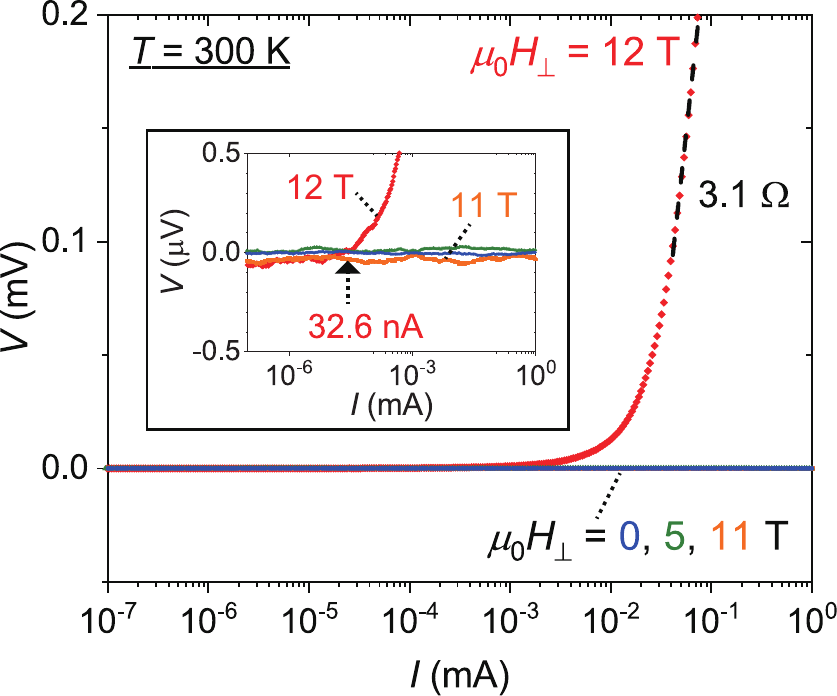}
\caption{\label{fig3}
\emph{V}--\emph{I} results at 300 K in various perpendicular magnetic fields. Inset, zoomed-in plot.
}
\end{figure}

The critical magnetic field of the RTSC was investigated in the PPMS at 300 K. Figure~\ref{fig3} shows the \emph{V}--\emph{I} characteristics of the RTSC in perpendicular magnetic fields of $\mu_{0}H_{\perp}$ = 0, 5, 11, and 12 T. For $\mu_{0}H_{\perp}\leq$ 11 T, the voltage does not increase for an applied current lower than 1 mA. At $\mu_{0}H_{\perp}$ = 12 T, however, a finite voltage is observed for currents exceeding the small value of 32.6 nA as shown in the inset. Thus, the critical magnetic field of the RTSC is approximately 12 T.

Here, an RTSC sample of the same design, but different to that used in Fig.~\ref{fig2} was used for this measurement [details of the \emph{R}--\emph{T} cycles associated with making this RTSC can be found in Fig. S2 in the \hyperref[supplmater]{supplementary material}]. Thus, the RTSC in this study is reproducible.

\subsection{Critical current of RTSC}
\label{sec:cricur}
\begin{figure}[h]
\includegraphics[width=8.5cm]{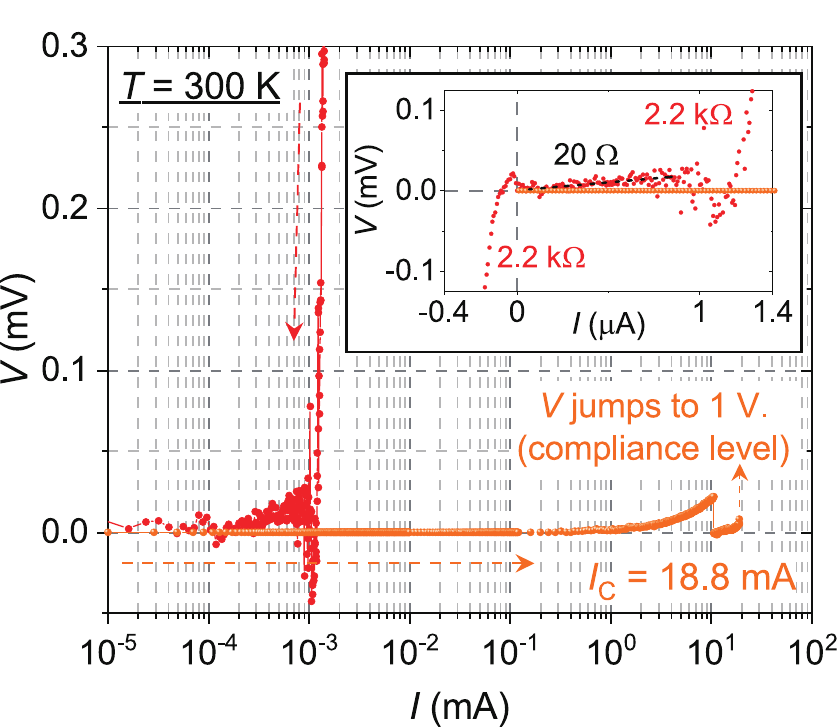}
\caption{\label{fig4}
Measurement of the DC critical current $I_{C}$ of the RTSC at 300 K in zero magnetic field; first, an applied direct current is increased in the positive direction (orange); after \emph{V} jumps to the set compliance level, \emph{V}--\emph{I} is measured again (red). Inset, \emph{V}--\emph{I} results in linear scale.
}
\end{figure}

The DC critical current $I_{C}$ of the RTSC was investigated in the PPMS at 300 K in zero magnetic field. The RTSC sample used for this measurement was that obtained after completing the \emph{R}--\emph{T} cycles in Fig.~\ref{fig2}. As shown in Fig.~\ref{fig4} (orange), by increasing the DC current from +0 mA, the voltage starts to slightly increase for currents exceeding +1 mA, which might be a thermal voltage due to Joule-heated aluminum bonded wires. Finally, at +18.8 mA, the voltage jumps to 1 V that is the set compliance level. Thus, the $I_{C}$ of the RTSC is 18.8 mA.

Here, $I_{C}$, normal resistance $R_{n}$ = 22.5 $\si{\ohm}$ at 300 K [purple, Fig.~\ref{fig2}], and the RTSC gap $\mathnormal{\Delta}$ = 0.8 eV (which will be shown later), satisfy the following Ambegaokar--Baratoff formula at 300 K:
\begin{equation}
I_{C}=\frac{\mathnormal{\Delta}}{2eR_{n}}\tanh\left( \frac{\mathnormal{\Delta}}{2k_{B}T} \right),\label{eq:ambeg}
\end{equation}
which gives $I_{C}$ of a JJ~\cite{Jos1962,Amb1963}. Thus, our consideration of the nature of the RTSC, i.e., as a capacitive Mott-insulating square lattice---essentially the square array of JJs, is justified.

\begin{SCfigure}[0.8][h]
\includegraphics[width=4.0cm]{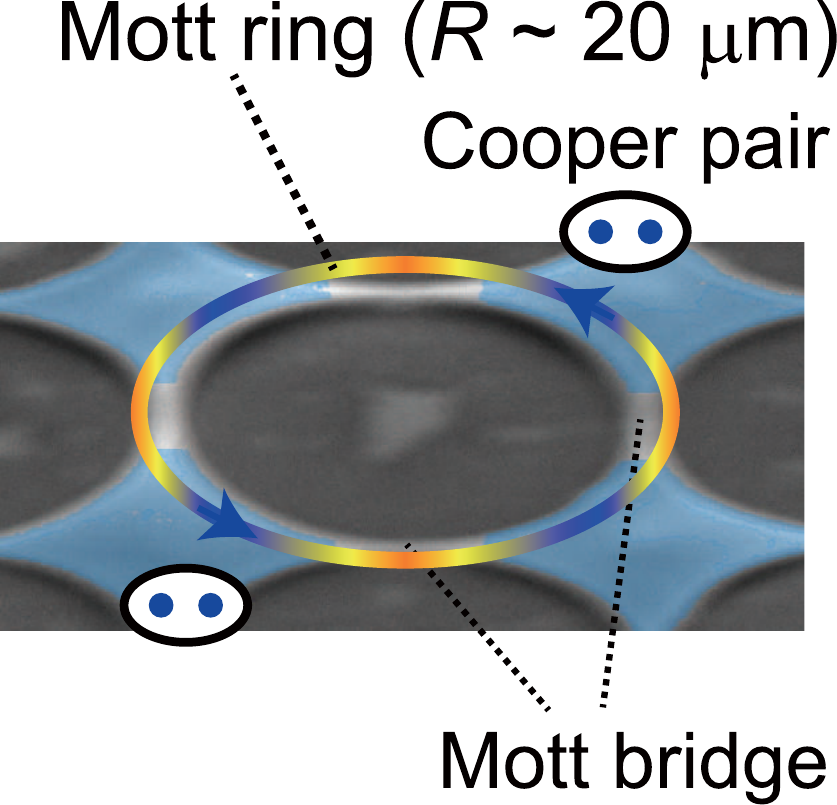}
\caption{\label{fig5}
Principle of the Mott ring; a circulating supercurrent path and Cooper pairs (twin blue dots) are illustrated using a false-color SEM.
}
\end{SCfigure}

Focusing on its unit cell, its schema is represented in Fig.~\ref{fig5}. Four Mott bridges form a 2D square closed circuit. Note that the so-called Mott ring is identical to the model element for Tinkham and \u{S}im\'{a}nek in considering JJ arrays [Section 6.6 in Ref.~\cite{Tin2004}] or granular SCs [Section 3.4 in Ref.~\cite{Sim1994}]. Under the principle, the periodic thru-holes are not a mere defect. Rather, it provides center-of-mass motion of the Cooper pair. More physically, the coherence length $\xi$ of the RTSC corresponds to the lattice constant of the artificial HbC, \emph{a} of 20 $\si{\micro}$m. The validity of the extraordinarily long $\xi$ is shown in the next section.

\vspace{-1.0cm}

\subsection{Advent of a new condensed phase}
\label{sec:scrq}

\begin{figure}[b]
\includegraphics[width=6.5cm]{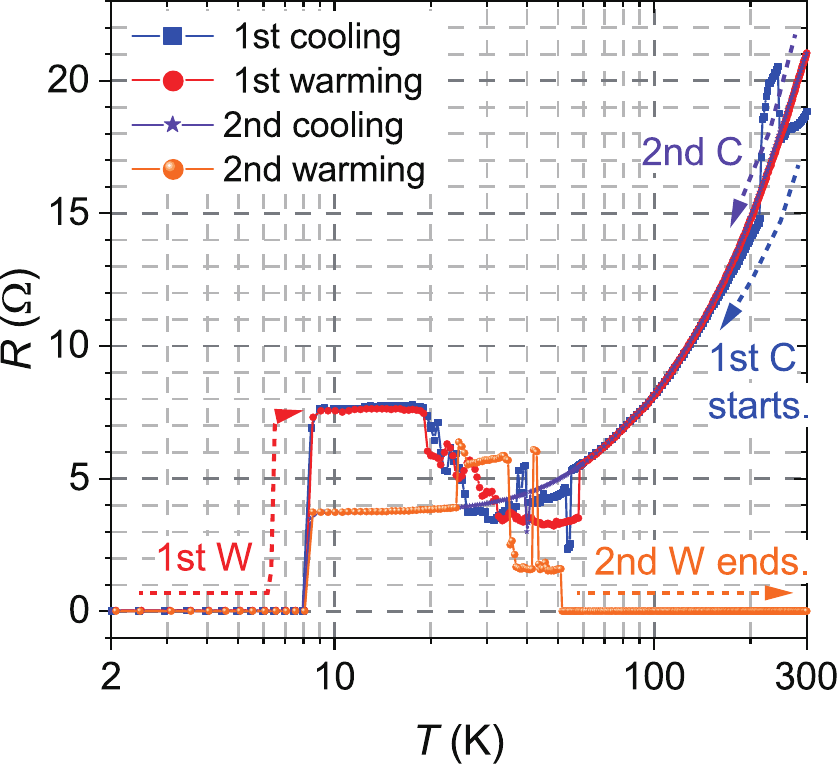}
\caption{\label{fig6}
Additional \emph{R}--\emph{T} cycles of the sample after its $I_{C}$ was measured in Fig.~\ref{fig4}; that is, before the first cooling, the sample was in the indeterminate state (i.e., the asymmetric mixture of metallic and insulating states). The second warming process (orange) realizes zero resistance at 300 K again; the new condensed phase is named the superconductor requiem (SCRQ).
}
\end{figure}

\begin{figure*}[tb]
\includegraphics[width=15.6cm]{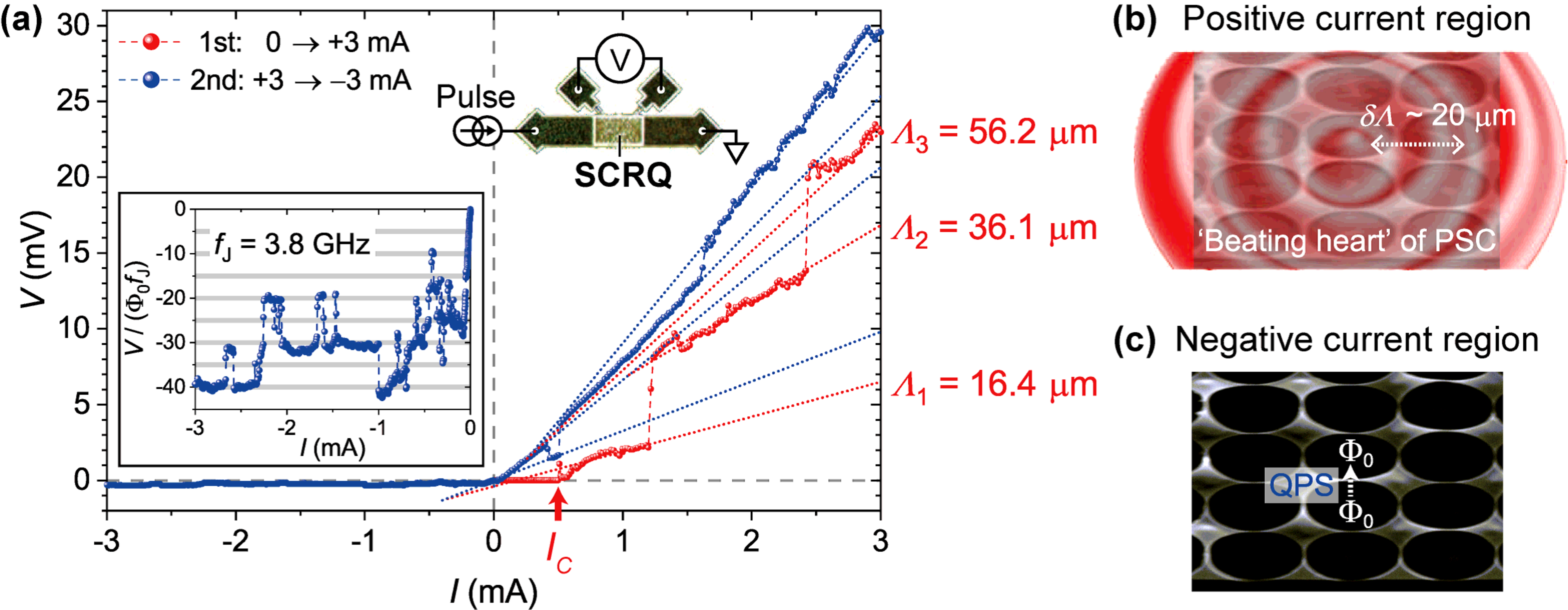}
\caption{\label{fig7}
(\textbf{a}) \emph{V}--\emph{I} results of the SCRQ at 300K. As shown in the circuit, a pulsed current is applied; first, the pulsed current is swept from +0 to +3 mA (red); second, from +3 to –3 mA (blue). The $I_{C}$ of the SCRQ is indicated (red arrow: 0.5 mA), and the diffusion length $\mathnormal{\Lambda}$ of nonequilibrium quasiparticles in the vicinity of the phase slip center (PSC) are denoted for each voltage step. Inset, zoomed-in plot of the negative current region; \emph{V} is normalized by the product of the flux quantum $\Phi_{0}$ and the Josephson plasma frequency $f_{J}$. (\textbf{b} and \textbf{c}) Schematic representation of the beating heart of the PSC in the positive current region and the quantum phase slip (QPS) in the negative current region, respectively, using false-color SEMs.
}
\end{figure*}

Immediately after the DC current exceeding $I_{C}$ was applied to the RTSC [orange, Fig.~\ref{fig4}], the \emph{V}--\emph{I} characteristics was measured again [red, Fig.~\ref{fig4}]. The result on a linear scale is shown in the inset. In terms of the resistance, the RTSC state disappears. Depending on the amount of the applied current, a metallic state of 20 $\si{\ohm}$ or an insulating state of 2.2 k$\si{\ohm}$ is realized. The anomalous point is that the \emph{V}--\emph{I} curve is the origin asymmetry; that is, the specimen is in an insulating state for a negative current, and, in contrast, it is in a mixed state of a metal and an insulator for a positive current. The positive-direction current is the same as that of the DC current used to investigate $I_{C}$. Excessively large current may force out some \emph{d} electrons filled in Mott-insulating bridges. In other words, holes are excessively injected into the RTSC.

Once a current that is larger than $I_{C}$ is applied, the original RTSC state disappears. Because there is nothing we can do about it, we continued investigation into the unusual state. First, the unusual sample was cooled and warmed again; its \emph{R}--\emph{T} characteristics are shown in Fig.~\ref{fig6}. During the second warming process (orange), the resistance decreased to zero at 50 K, and the zero-resistance state remains up to 300 K. However, the sample is no longer a simple RTSC. For convenience, we name the new condensed phase the superconductor requiem (SCRQ).

Immediately after the \emph{R}--\emph{T} procedure, the \emph{V}--\emph{I} characteristics were investigated in the PPMS at 300 K in a near-zero 3.6 $\si{\micro}$T perpendicular magnetic field. As shown in the circuit in the inset of Fig.~\ref{fig7}(a), a pulsed current was applied to the SCRQ, and the voltage was measured using the four-probe method [see ``Materials and Methods'' in the \hyperref[supplmater]{supplementary material} for details]. The first current sweep toward the positive direction (red), from +0 to +3 mA, confirms a finite voltage for currents exceeding 0.5 mA (i.e., $I_{C}$ of the SCRQ is 0.5 mA) and the subsequent several voltage steps. This phenomenon has been confirmed in previous studies on an SC crystal whisker and a microbridge made from an SC thin film and is the nonequilibrium SC phenomenon above $I_{C}$, known as the \emph{beating heart of the phase slip center (PSC)} [Section 11.6 in Ref.~\cite{Tin2004} and Ref.~\cite{Mey1972}]. Where, PS signifies that the phase of the SC order parameter $\mathnormal{\Delta}$ increases at different rates on either side of the localized resistive center. The diffusion length $\mathnormal{\Lambda}$ of nonequilibrium quasiparticles in the vicinity of the beating heart of the PSC is expressed as follows~\cite{Tin2004}:
\begin{equation}
\mathnormal{\Lambda}=V\times l/\{ 2R_{n}(I-I_{C}/2) \},\label{eq:beating}
\end{equation}
where \emph{l} is the distance between the voltage pads (0.3 mm). By substituting the normal resistance $R_{n}$ = 21 $\si{\ohm}$ at 300 K [purple, Fig.~\ref{fig6}] and 0.5 mA for $I_{C}$ [red arrow, Fig.~\ref{fig7}(a)], $\mathnormal{\Lambda}$ values of 16.4 $\si{\micro}$m, 36.1 $\si{\micro}$m, and 56.2 $\si{\micro}$m, respectively, are obtained for each voltage step. Of note, the differences between subsequent steps are $\sim$20 $\si{\micro}$m; that is, the PS center expands its region stepwise with an interval $\delta\mathnormal{\Lambda}$ $\sim$ 20 $\si{\micro}$m, as schematically shown in Fig.~\ref{fig7}(b), which corresponds to \emph{a} of the artificial HbC. Tinkham assumed that PS occurs in the core length of $\xi$, i.e., the coherence length that governs the spatial variation of $\mathnormal{\Delta}$~\cite{Tin2004}. Thus, the magnitude of $\xi$ of the RTSC ($\sim$20 $\si{\micro}$m), which has already been estimated based on the principle of the Mott ring [Fig.~\ref{fig5}], is verified by this PS phenomenon.

Subsequently, the pulsed current was swept from +3 to –3 mA [blue, Fig.~\ref{fig7}(a)]. Although the temporal interval required to start this measurement was as short as 52 seconds, the voltage at +3 mA was already larger than that at the end of the first current sweep (red). Several voltage steps are also confirmed with a decrease in the current toward zero. Thus, PS occurs in the positive current region. In contrast, in the negative current region, a voltage was not generated upon cursory inspection. However, a small, discretized voltage was generated as shown in the inset of Fig.~\ref{fig7}(a). The voltage is normalized by the product of the flux quantum $\Phi_{0}$ and the abruptly introduced Josephson plasma frequency $f_{J}$ = 3.8 GHz. The Lorentz force, which is a vector product of the applied current and the magnetic flux density of $\Phi_{0}$, drives $\Phi_{0}$ itself as schematically shown in Fig.~\ref{fig7}(c); the movement of $\Phi_{0}$ generates discretized, resistive voltage. This is the phenomenon that is exactly dual to Josephson tunneling: the quantum phase slip (QPS)~\cite{Moo2006,Ras2013,Erg2013,Erg2017}. The magnetic field required to create a single $\Phi_{0}$ in a 19.8-$\si{\micro}$m-diameter thru-hole of the SCRQ is 6.7 $\si{\micro}$T. Although the measurement is performed at $\mu_{0}H_{\perp}$ = 3.6 $\si{\micro}$T, 1 mA, for instance, passing through the vicinity of the thru-hole creates 20 $\si{\micro}$T at the center of the thru-hole; therefore, $\Phi_{0}$ can exist. Indeed, a larger current results in a clearer discretization of voltage [inset, Fig.~\ref{fig7}(a)].

In conclusion, the SCRQ exhibits PS and QPS in the positive and negative current regions, respectively, at 300 K. Both are SC-related phenomena; that is, the SCRQ still contains the character of an RTSC. In the former part of this section, we assumed that the excessively large current in the positive direction injected excessive holes into the RTSC, which was the precursor of the SCRQ. Co-existence of the SC and hole-enriched states results in a considerable consequence.

\subsection{P-type semiconducting--superconducting interface in the SCRQ}
\label{sec:psemisc}
Hereafter, all experimental data were obtained in the atmosphere, i.e., at room temperature, in a terrestrial magnetic field, and under atmospheric pressure.

\begin{SCfigure*}[0.5][t]
\includegraphics[width=12cm]{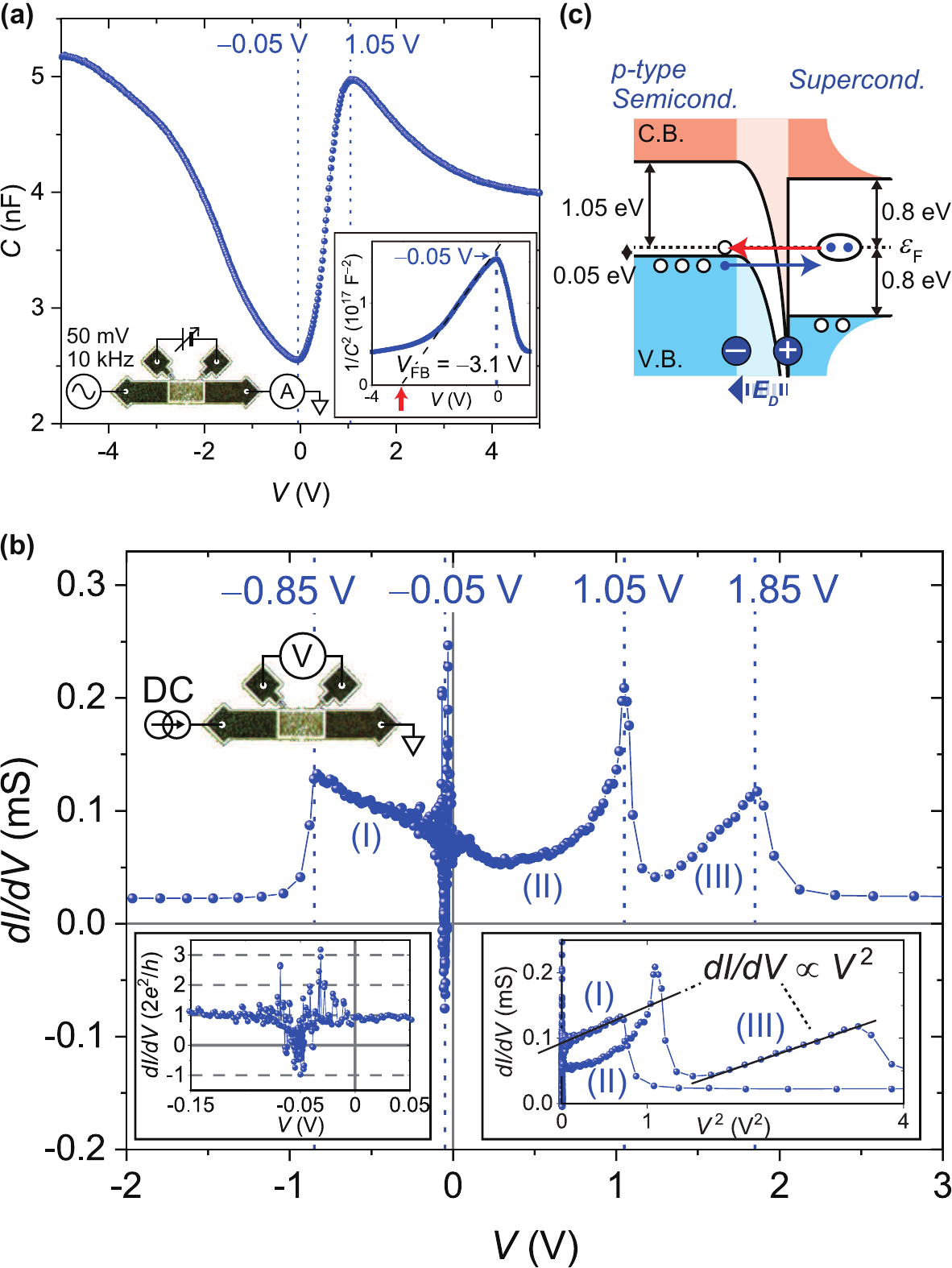}
\caption{\label{fig8}
(\textbf{a}) \emph{C}--\emph{V} results of the SCRQ in the atmosphere. As shown in the circuit, for each biased DC voltage, \emph{C} is determined by temporally integrating the measured current. Characteristic voltages are denoted. Inset, Mott--Schottky plot, i.e., $1/C^{2}$ \emph{vs.} \emph{V} plot; the deep flat-band potential $V_{FB}$ is indicated (red arrow). (\textbf{b}) $dI/dV$--\emph{V} results of the SCRQ in the atmosphere. As shown in the circuit, a DC current is applied, and the voltage is measured. The $dI/dV$ is obtained by differentiation. Characteristic voltages are denoted. (Bottom-left inset) Zoomed-in plot near the origin; $dI/dV$ is normalized by the conductance quantum, $G_{0} = 2e^{2}/h$. (Bottom-right inset) $V^{2}$-dependent $dI/dV$; the voltage regions denoted as (I)--(III) correspond to those in the main panel. (\textbf{c}) Constructed semiconductor model by unifying all experimental results in (a) and (b); that is, a special Schottky junction consisting of the hole-enriched semiconductor (p-Semi) and SC is formed in the SCRQ; left side, p-Semi with the band gap $E_{g}$ = 1.1 eV; right side, SC with the gap $\mathnormal{\Delta}$ = 0.8 eV. Shades: light blue, valence band (V.B.); light orange, conduction band (C.B.). The Fermi level of p-Semi is assumed to be 0.05-eV above the edge level of its V.B.; shown here is the case of zero-bias voltage, i.e., which is equivalent to that of SC, $\varepsilon_{F}$. Dots: empty, a hole; blue, an electron; twin, Cooper pair. The Cooper pair is confined in the deep potential well because of the large $V_{FB}$. Solid arrows: transport of a hole (red) and an electron (blue). Broken arrow, the built-in electric field $E_{D}$; direction, polarity. For clarity, the density of states of the \emph{d}-wave SC is designedly changed to that of an \emph{s}-wave SC, and thermally excited carriers are disregarded because $\mathnormal{\Delta}\gg k_{B}T$ at room temperature.
}
\end{SCfigure*}

Figure~\ref{fig8}(a) shows the \emph{V}-dependent capacitance \emph{C} of the SCRQ. As shown in the circuit in the inset, for each biased DC voltage \emph{V}, the accumulated charge in the SCRQ was excited by an AC voltage with an amplitude of $\pm$50 mV and a frequency of 10 kHz, and the amount of charge was obtained by temporally integrating the measured current; consequently, \emph{C} was determined for each DC voltage \emph{V}. The \emph{C}--\emph{V} curve exhibits the characteristics of a p-type Schottky diode, in which the depletion layer and the inversion layer are formed in the ranges of \emph{V} $<$ –0.05 V and –0.05 $<$ \emph{V} $<$ +1.05 V, respectively. Moreover, the Mott--Schottky plot (i.e., $1/C^{2}$ \emph{vs.} \emph{V} curve) shown in the inset, that is obtained from the main panel, confirms the existence of a deep flat-band potential $V_{FB}$ of –3.1 V in the SCRQ.

Figure~\ref{fig8}(b) shows the differential conductance $dI/dV$ \emph{vs.} \emph{V} result of the SCRQ. As shown in the circuit in the inset, a DC current was applied to the SCRQ, and the voltage was measured using the four-probe method. The $dI/dV$ was obtained by differentiation. Due to the physically meaningful, large capacitance \emph{C} of several nF of the SCRQ [Fig.~\ref{fig8}(a)], the DC electrical property is considerably different from that of Fig.~\ref{fig7}(a) which was obtained by applying a pulsed current. That is, a pulsed current flows through the SCRQ easily because it is not impeded by \emph{C}, whereas \emph{C} has an effect on the DC current. Four peaks at \emph{V} = –0.85, –0.05, +1.05, and +1.85 V are clearly confirmed. Note that two of them correspond to \emph{V} of the observed \emph{C}--\emph{V} extremums in Fig.~\ref{fig8}(a), which implies that p-type semiconducting (p-Semi) carriers, i.e., holes, take part in the electrical conductance.

The $dI/dV$ characteristics near the origin are enlarged in the inset at the bottom-left corner of Fig.~\ref{fig8}(b). The $dI/dV$ is normalized by the conductance quantum, $G_{0} = 2e^{2}/h$, where $h$ is Planck’s constant. Clearly, the SCRQ has an ideal quantum regime (i.e., $dI/dV = G_{0}$). Moreover, the suppression of $G_{0}$ in the vicinity of \emph{V} = –0.05 V indicates the occurrence of QPS. Furthermore, the $dI/dV$ does not merely drop to zero conductance, but rather, it undergoes a reversal to $-G_{0}$. The fluctuating sign of $G_{0}$, which is observed in the vicinity of QPS, supports the existence of charge-neutral Majorana fermions~\cite{Bee2012,Hec2012,Vij2016,Mou2012}. The details are further discussed in the \hyperref[sec:iiia]{Section III. A}.

The inset at the bottom-right corner in Fig.~\ref{fig8}(b) is the $V^{2}$-dependent $dI/dV$ that is obtained from the main panel. In the voltage regions ranging from –0.85 to –0.05 V and from +1.05 to +1.85 V, denoted as (I) and (III) in the main panel, respectively, the $dI/dV$ is proportional to $V^{2}$. The SC gap which satisfies this relation has a \emph{d}-wave symmetry~\cite{Mon1992,Tan1996}. Thus, superconductivity in this study is due to \emph{d} electrons, which is a natural consequence of the artificial HbC---the precursor of the RTSC and SCRQ---consisting of \emph{d}-orbital-filled Mott insulators. Moreover, the widths of regions (I) and (III) are both 0.8 V; thus, the \emph{d}-wave SC gap $\mathnormal{\Delta}$ is 0.8 eV. The value of $\mathnormal{\Delta}$ has already been justified by Eq.~(\ref{eq:ambeg}) when estimating the $I_{C}$ of the RTSC.

The residual region, ranging from –0.05 to +1.05 V, denoted as (II) in Fig.~\ref{fig8}(b), is regarded as a forcibly opened gap that is created by the excessive injection of holes [inset, Fig.~\ref{fig4}]. That is, we assume that therein lies the p-Semi order with a band gap $E_{g}$ of 1.1 eV. In conclusion, the p-Semi and SC states coexist in the SCRQ; thus, a special Schottky junction is formed in the SCRQ. By unifying the above results, the semiconductor model [Section 3.8 in Ref.~\cite{Tin2004}] is assembled in Fig.~\ref{fig8}(c), in which p-Semi with $E_{g}$ = 1.1 eV and SC with $\mathnormal{\Delta}$ = 0.8 eV are shown on the left and right sides, respectively. The twin electron (blue dots) is the Cooper pair. The Fermi level (black dotted line) of the p-Semi is assumed to be 0.05-eV higher than the edge level of its valence band (V.B.), so that every feature of the \emph{C}--\emph{V} [Fig.~\ref{fig8}(a)] and $dI/dV$--\emph{V} results [Fig.~\ref{fig8}(b)] is completely included in the model. The case of zero-bias voltage (i.e., the Fermi level of the p-Semi and that of the SC, $\varepsilon_{F}$, are equivalent) is shown. See Fig. S3 in the \hyperref[supplmater]{supplementary material} for all other non-zero-bias regions, that are explained by setting four distinct voltages (i.e., \emph{V} = –0.85, –0.05, +1.05, and +1.85 V) as electrical potential boundaries from the basis of $\varepsilon_{F}$. Because $V_{FB}$ = –3.1 V, the Cooper pair is confined in a deep potential well. For clarity, the density of states of the \emph{d}-wave SC is designedly changed to that of an \emph{s}-wave SC. In addition, thermally excited carriers are disregarded because $\mathnormal{\Delta}\gg k_{B}T$ at room temperature.

At the interface of the p-Semi--SC junction, the diffusion of holes from the p-Semi to the SC is promoted because of the nonequilibrium carrier concentration, as is the case for a conventional Schottky junction. Consequently, the diffusion creates a built-in electric field $E_{D}$, the polarity of which is denoted by the broken arrow. However, the subsequent carrier transport is quite different from that of a conventional Schottky junction. In the case of the p-Semi--SC junction [see Fig.~\ref{fig8}(c)], an electron that has just been forced out from the p-Semi to SC by $E_{D}$ (blue arrow) is Andreev-reflected back into the p-Semi (red arrow) as a hole (empty dot), because an available density of states (DOS) is empty in the SC gap region [Section 11.5 in Ref.~\cite{Tin2004} and Refs.~\cite{And1964,Blo1982}]. The probability of the Andreev reflection (AR) is highest when the edge of the electron-rich V.B. of the p-Semi corresponds to the $\varepsilon_{F}$ of the SC, i.e., at \emph{V} = –0.05 V, at which the sign of $G_{0}$ is inverted [see the inset at the bottom-left corner of Fig.~\ref{fig8}(b)]. Thus, in the vicinity of the QPS center, the electron and the hole bind to each other in compliance with the law of causality, forming a zero-net-charge Majorana fermion~\cite{Bee2012}.

The most important consequence of the p-Semi--SC junction model is that the AR-assisted process never reaches thermal equilibrium [see Fig.~\ref{fig8}(c)]. Because the return of the AR-assisted hole to the p-Semi side does not dissolve the initial nonequilibrium carrier concentration at all, the hole diffuses again to create $E_{D}$, which promotes the subsequent injection of an electron into the SC and the consequent return of an AR-assisted hole to the p-Semi side, again. Thus, as long as the carrier diffusion is accompanied by AR, thermal equilibrium is never achieved, which ensures the sustainability of the junction system. By taking into account that the opposed flow of oppositely charged carriers creates a net current, the SCRQ will generate a sustainable current.

\subsection{Current arising from SCRQ}
\label{sec:lowloss}

\begin{figure*}[tb]
\includegraphics[width=15.6cm]{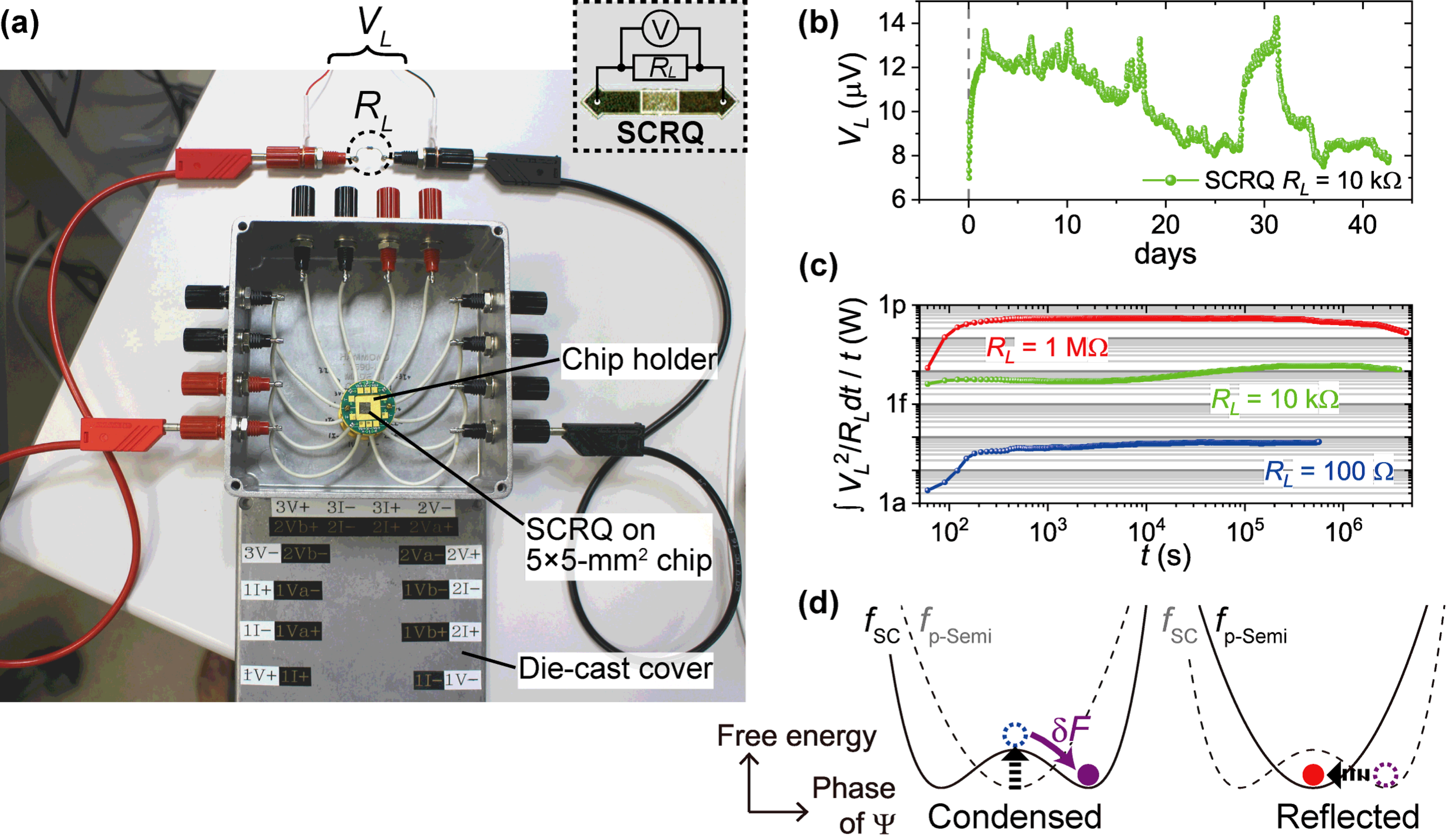}
\caption{\label{fig9}
(\textbf{a}) Photograph of the experimental setup for the discharge test. During measurements, the SCRQ is enclosed in a die-cast box to prevent Vis-light irradiation. $R_{L}$ is a commercial metal film resistor, and the voltage, $V_{L}$, across $R_{L}$ is measured continuously using a nanovoltmeter. Inset, equivalent circuit for the discharge test configuration. (\textbf{b}) Time evolution of $V_{L}$ for $R_{L}$ = 10 k$\si{\ohm}$ in the atmosphere. (\textbf{c}) Time evolution of the time-averaged consumed energy for several $R_{L}$. (\textbf{d}) Schematic representation of the sustainable process using the Ginzburg--Landau free energy function \emph{f}, where the horizontal axis is the phase of the macro wavefunction $\Psi$; \emph{f}\textsubscript{p-Semi}, \emph{f} in p-Semi; \emph{f}\textsubscript{SC}, \emph{f} in SC. $\delta$\emph{F}, the condensation energy. Circles, charged particles.
}
\end{figure*}

To examine whether the SCRQ generates a current and whether it is sustainable or not, a simple experiment was performed. Figure~\ref{fig9}(a) shows a photograph of the experimental setup, and its equivalent circuit is shown in the inset at the upper-right corner. The load resistor $R_{L}$ (a commercial metal film resistor) was serially connected to the SCRQ, and the voltage $V_{L}$ across $R_{L}$ was chronologically measured using a nanovoltmeter; performing what is referred to as the discharge test in the field of batteries. The discharge test was executed with the SCRQ enclosed in a die-cast box in order to prevent Vis-light irradiation from perturbing the p-Semi--SC junction. As discussed in the next section, the Vis-light photon energy, which is equivalent to $E_{g}$ (1.1 eV) and 2$\mathnormal{\Delta}$ (1.6 eV), affects the electrical properties of the SCRQ. Thus, the SCRQ must be characterized in the dark. Otherwise, its greatest feature cannot be distinguished from that of a p--n junction-type solar cell.

The discharge tests were executed using $R_{L}$ = 100 $\si{\ohm}$, 10 k$\si{\ohm}$, and 1 M$\si{\ohm}$. Figure~\ref{fig9}(b) shows the case for $R_{L}$ = 10 k$\si{\ohm}$ [all other cases can be found in Fig. S4 in the \hyperref[supplmater]{supplementary material}]. Although the result is strongly dependent on the magnitude of $R_{L}$, current certainly flows into the external circuit. Thus, the SCRQ can drive electronic components, just like a battery. The median value of the AR-assisted current $I_{A}=V_{L}/R_{L}$ for $R_{L}$ = 10 k$\si{\ohm}$ during the measurement period of 43 days is 1.1 nA.

Figure~\ref{fig9}(c) shows the time-averaged consumed energy for each $R_{L}$, which is expressed as $\overline{P_{L}}=\int V_{L}^{2}/R_{L}dt\big/ t$, and it indicates that the SCRQ produces at most 1 pW. For $R_{L}$ = 10 k$\si{\ohm}$ (green) and 1 M$\si{\ohm}$ (red), the produced energy tends to decrease over time, this effect being most pronounced for the case of 1 M$\si{\ohm}$. Nevertheless, the several confirmed peaks in Fig.~\ref{fig9}(b) signify the recovery of the current against the deterioration, which cannot be understood by analogy to the mechanism of conventional batteries in which charge diffusion dominates the system.

\section{DISCUSSION}
\label{sec:discussion}

\subsection{Combined Josephson and charging effects in the SCRQ: Two distinct regimes}
\label{sec:iiia}
Consideration of the Josephson coupling energy $E_{J}$ and the charging energy $E_{C}$ is crucial for further inquiry into the SCRQ that consists of JJ arrays. $E_{C}$ is defined as the energy that is necessary to charge a single JJ with \emph{e} [Section 7.3 in Ref.~\cite{Tin2004} and Refs.~\cite{Ave1985,Ful1989}]. As each JJ is characterized by the SC gap $\mathnormal{\Delta}$ (0.8 eV), $E_{C}$ is calculated as $E_{C}=e^{2}/2C_{0}=e^{2}/2(e^{2}/\mathnormal{\Delta})=\mathnormal{\Delta}/2$ = 0.4 eV, where $C_{0}$ is the capacitance of a single JJ. Thus, $E_{C}$ is much larger than $k_{B}T$ at room temperature, which ensures the charging effect in the atmosphere. The validity of $E_{C}$ was also confirmed by measuring the Vis-spectral response of the SCRQ for the photon wavelength $\lambda$ range 300--700 nm [Fig. S5, \hyperref[supplmater]{supplementary material}]. The observed concave peak at $\lambda$ $\sim$ 400 nm (i.e., photon energy, $h\nu$, $\sim$3.1 eV) indicates that the photo current is suppressed in this region of $\lambda$. That is, $E_{g}$ (1.1 eV) and 2$\mathnormal{\Delta}$ (1.6 eV) essentially disappear, and the rest of the irradiating photon energy, i.e., $h\nu-(E_{g}+2\mathnormal{\Delta})$ = 0.4 eV, is consumed as $E_{C}$.

On the other hand, $E_{J}$ is given by $E_{J}=I_{C}\hbar/2e$, where $\hbar$ is Dirac’s constant~\cite{Ful1989,Moo2006}. By substituting the $I_{C}$ of the SCRQ, i.e., 0.5 mA [red arrow, Fig.~\ref{fig7}(a)], we determine that $E_{J}$ = 1 eV, which indeed satisfies the special condition of $E_{J}\gtrsim E_{C}$. In the regime, Heck {\it et al.}~\cite{Hec2012} and Vijay {\it et al.}~\cite{Vij2016} have proposed various types of JJ arrays to control interior Majorana fermions. Beenakker shows a generic route to Majorana fermions [Section II in Ref.~\cite{Bee2012}], and what we have done in creating the SCRQ, i.e., injecting asymmetry into the RTSC [inset, Fig.~\ref{fig4}] and subsequent re-opening $\mathnormal{\Delta}$ [Fig.~\ref{fig6}], just follow the pathway. In this study, the Majorana bound state~\cite{Bee2012,Mou2012} is realized by confinement of Cooper pairs in the deep quantum well ($V_{FB}$ = –3.1 V) with the help of AR [Fig.~\ref{fig8}(c)]. Specifically, in the vicinity of the QPS center (i.e., \emph{V} = –0.05 V), the $|dI/dV|$ equals to a quantized amount, $G_{0}$ of $2e^{2}/h$ [bottom-left inset, Fig.~\ref{fig8}(b)]. Where, the factor of two is due to the fact that AR of an electron into a hole doubles the current, and equality signifies that the AR probability is unity~\cite{Bee2012}. The probability unity is distinctly different from the quantum mechanical probability theory of which effect is obtained by taking all possible causal wavefunctions into account and is therefore probabilistic. The probability unity, by contrast, provides the Newtonian deterministic future, and we no longer have to take care about the infinitely many historical paths (whether or not they really exist). When such paths are completely absent, only two points, i.e., a cause and its effect, remain in the space--time coordinate. That is, no matter how fast or slow time goes by, an electron in the past definitely encounters a hole in the future. The decisive causality, which may be also referred to as ``superoccasion (at the same position)'' in contrast to the superposition principle of quantum mechanics, realizes a zero-net-charge Majorana.

At the macroscopic level, a cause precedes its effect in general, which is mostly due to the second law of thermodynamics: there is an obvious direction of the time’s arrow (although we do not know why it is so). At the p-Semi--SC interface, on the other hand, the cause and the effect are indistinguishable. Under time reversal, the eventual Andreev-reflected hole in the future is reversely injected into SC. Here, we are no longer able to distinguish whether the hole is going toward the past or merely diffuses in the ordinal direction of the time’s arrow (i.e., toward the future leading to AR of an electron) because of the nonequilibrium charge concentration at the interface. Thus, the entire p-Semi--SC junction system is time-symmetric (even though the system includes the diffusion process which follows the second law of thermodynamics). In other words, the abovementioned decisive causality that binds an electron in the past (cause) to a hole in the future (effect) cannot keep the one-way property: the time’s arrow is reversible. This is why the $dI/dV$ undergoes a reversal to $-G_{0}$ at the QPS center [bottom-left inset, Fig.~\ref{fig8}(b)]. The real Majorana sometimes runs toward the past, which should be the nature of a particle--antiparticle identical Majorana.

In contrast, under intense Vis-light illumination, the SCRQ undergoes a complete change, entering a distinct regime. Such a Vis-light photon easily excites a carrier from the V.B. to the C.B., and the excited carriers that jump across $E_{g}$ and/or 2$\mathnormal{\Delta}$ participate in the tunneling process. As shown in Fig. S6(a) in the \hyperref[supplmater]{supplementary material}, the transport current of the DC-voltage-biased SCRQ is increased under illumination. Its normalized \emph{I}--\emph{V} curve in Fig. S6(b) using $\mathnormal{\Delta}$ (0.8 eV) and $R_{n}$ (181.1 k$\si{\ohm}$), in accordance with the procedure of Blonder {\it et al.}~\cite{Blo1982}, proves that the p-Semi--SC interface is sufficiently barrier-free to observe carrier tunneling. In this regime, $E_{J}$ is given by $E_{J}=(R_{Q}⁄R_{n})\times \mathnormal{\Delta}/2$ = 14 meV, where $R_{Q}$ is the quantum resistance (6.45 k$\si{\ohm}$)~\cite{Erg2013,Hav1994,Fit1998}. In this condition of $E_{J}\ll E_{C}< \mathnormal{\Delta}$, the SCRQ shows the envelope~\cite{Her1994} of the \emph{I}--\emph{V} curves [Fig. S6(a)] and periodic oscillation~\cite{Hav1994} of the $dI/dV$ curves [Fig. S6(c)] while sweeping the voltage, which is the Coulomb blockade (CB) of 2\emph{e}--\emph{e} parity effect~\cite{Hav1994,Fit1998,Her1994,Eil1994,Ama1994,Pek2008}.

In general, a double-JJ configuration, which capacitively couples with a gate electrode, is necessary to observe CB. Otherwise, the charging effect fades out due to directly connected low-impedance leads [Section 7.2 in Ref.~\cite{Tin2004}]. However, this is not the case for the SCRQ. By simplifying the SCRQ as the RCSJ model, in which the SCRQ is shunted by $R_{n}$ (181.1 k$\si{\ohm}$) and the total capacitance \emph{C}, the time constant of the circuit is calculated as $CR_{n}$ $\sim$ 10\textsuperscript{–4} s, where the \emph{C}--\emph{V} result [Fig.~\ref{fig8}(a)] near the origin is used as \emph{C} (2.6 nF). The long time constant ensures that the SCRQ can see the high-impedance voltage source sufficiently well, and therefore, the directly connected leads do not disturb the observation of CB. In other words, the directly connected two-probe DC bias source easily charges the elementary JJ of the SCRQ. This is why sweeping the voltage results in the formation of the \emph{I}--\emph{V} envelope and the $dI/dV$ oscillation [details can be referred from Figs. S6 and S7 in the \hyperref[supplmater]{supplementary material}], and it is also true for the SCRQ in the QPS regime. Actually, the \emph{C}--\emph{V} [Fig.~\ref{fig8}(a)] and $dI/dV$ curves [Fig.~\ref{fig8}(b)] wobble while sweeping the voltage and current, respectively [details can be referred from Fig. S8 in the \hyperref[supplmater]{supplementary material}]. The details of this behavior are described therein in connection with another nonequilibrium SC phenomenon known as the \emph{enhancement of $\mathnormal{\Delta}$ by the extraction of quasiparticles} [Section 11.3 in Ref.~\cite{Tin2004} and Ref.~\cite{Chi1979}].

In conclusion, the SCRQ, the nature of which has the QPS regime (i.e., $E_{J}\gtrsim E_{C}$), enters the CB regime (i.e., $E_{J}\ll E_{C}< \mathnormal{\Delta}$) under intense Vis-light illumination because $\mathnormal{\Delta}$ (0.8 eV) lies in the Vis region. This fact is also important for the RTSC. If one wants to maintain the RTSC state, the sample should be kept in the dark; nevertheless, complete darkness is unnecessary; for example, enclosure in an acrylic box wrapped in aluminum foil is satisfactory. The RTSC sample, which was used in the experiment for the critical magnetic field, had been kept in such an acrylic box for 28 days after its superconductivity was obtained by repeating \emph{R}--\emph{T} cycles [Fig. S2, \hyperref[supplmater]{supplementary material}], and retained the RTSC state as shown in Fig.~\ref{fig3}.

\subsection{Role of repeating \emph{R}--\emph{T} cycles in creating RTSC}
\label{sec:iiib}
The role of the initial part of the \emph{R}--\emph{T} cycles is to create the HbC, i.e., the 2D square lattice in which conductive Nb islands and Mott-insulating Nb bridges are alternately arranged [Fig.~\ref{fig1}]. By considering the above discussion of $E_{J}$ \emph{vs.} $E_{C}$, the latter part of the \emph{R}--\emph{T} cycles seems necessary to balance \emph{U} and \emph{t} in the artificial HbC. That is, the pulsed excitation current, which is utilized to measure the resistance during the \emph{R}--\emph{T} cycles, is capacitively supplied to the 2D Mott-insulating square lattice, and the consequent moderate balance of \emph{U} and \emph{t} results in the RTSC, just as optimal doping realizes HTSCs~\cite{Fuk2009,Sca1995}. In some cases, the \emph{R}--\emph{T} cycles resulted in a near-miss RTSC in which the resistance at 300 K did not completely drop to zero [see Fig. S9(a) in the \hyperref[supplmater]{supplementary material}]. Nevertheless, the subsequent current cycles at 300 K realized zero resistance for a certain current range [Fig. S9(b), \hyperref[supplmater]{supplementary material}], which indicated the importance of charging in creating RTSCs. More precisely, spatially homogeneous charging is necessary; otherwise, the disorderly charge-density distribution easily results in 2D Anderson localization, and the PnC sample is hastily forced to become an insulator as shown previously in Ref.~\cite{Zen2019}.

So far, the method to elucidate what mediates the 2D \emph{e}--\emph{e} interaction relies on elimination, because we cannot identify any candidates other than the forcibly created 2D BZ of PnC~\cite{Zen2019}. Direct proof of the mediating phonons will be provided by IR and/or THz spectroscopy during temperature cycling~\cite{Mak2019}, because the small $h\nu$ of these waves enable the monitoring of phonons without disturbing the advent of $\mathnormal{\Delta}$.

Still, the question of how the anomalous phonon-mediated \emph{e}--\emph{e} interaction can survive even after the PnC is warmed up to 300 K, where phonons are in the IR region and electrons are in the UV region, remains. A possible answer is a Peierls-type transition, in which the charge-modulated electron system conversely affects the PnC phonons, and the mutual reaction gradually brings their energy levels closer, by repeating temperature cycles. This scenario may agree with the consequent ground state that has the energy level of $\mathnormal{\Delta}$ (0.8 eV), lying between the IR and the UV regions. Also, it may explain the important requisites for the \emph{R}--\emph{T} procedure, i.e., the slow cooling--warming rate to observe abnormal phase transitions [e.g., see Table S1 in the \hyperref[supplmater]{supplementary material}]. For example, a pioneering organic superconductor of a tetra-methyl-tetra-selena-fulvalene (TMTSF) salt drives the Peierls instability at ambient pressure. Specifically, the ground state of (TMTSF)\textsubscript{2}ClO\textsubscript{4} depends on the cooling rate, i.e., superconductivity is achieved only when cooled slowly; otherwise, a spin-density-wave insulating ground state is exhibited because the rapid cooling process fails to order ClO\textsubscript{4}\textsuperscript{–} anions~\cite{Abe1988,Pou2018}. Investigating the possibility of the charge-induced Peierls instability in PnC, along with whether another aspect of the PnC---the reduced thermal conductance---is relevant in this scenario, requires multilateral analysis, such as, for example, THz spectroscopy, NMR, and XRD.

\subsection{Thermodynamic perspectives on the low-loss current}
\label{sec:iiic}
The low-loss current [Fig.~\ref{fig9}] does not violate the second law of thermodynamics. The sustainability originates from AR and carrier diffusion that bind to each other in compliance with the law of causality [see Fig.~\ref{fig8}(c)]. Thermodynamically, the p-Semi--SC interface which is in the QPS regime (i.e., in the dark) is a reversible system. Additionally, $E_{g}$ and 2$\mathnormal{\Delta}$, which are much larger than $k_{B}T$ at room temperature, ensure that the interface is a thermally isolated system. In the reversible, isolated system, the entropy \emph{S} never increases~\cite{Fer1937}, and the system never ends up in a silent, final state. Thus, the sustainability is protected by thermodynamics.

By using the Ginzburg--Landau free energy function \emph{f} [Chapter 4 in Ref.~\cite{Tin2004}], where the horizontal axis is the phase of the macro wavefunction $\Psi$, the sustainable process is illustrated in Fig.~\ref{fig9}(d). $E_{D}$ does the work for a charged particle, but at the moment of its entrance into the SC phase, the particle gains the condensation energy $\delta$\emph{F}. (The particle does not really enter the SC phase because an available DOS is empty. What gains $\delta$\emph{F} in the SC phase is its antiparticle~\cite{Dir1958}---precisely, Majorana~\cite{Bee2012}---as already discussed in the \hyperref[sec:iiia]{Section III. A}.) If $\delta$\emph{F} overcomes the work done by $E_{D}$, the particle is Andreev-reflected back into the p-Semi phase with its energy increased (along with its charge inverted), as if a ball bounced back with a coefficient of restitution greater than 1.

The net current, which flows into the external circuit, is approximately $I_{A}$ $\sim$ 1.1 nA [Fig.~\ref{fig9}(b)], and its origin of the 2\emph{e}-charge transport begins at a JJ somewhere in the SCRQ. Here, the Josephson plasma frequency $f_{J}=\sqrt{eI_{C}/\pi hC}$ [Section 6.3 in Ref.~\cite{Tin2004}] of the SCRQ in the QPS regime, where $I_{C}$ = 0.5 mA [red arrow, Fig.~\ref{fig7}(a)] and the total capacitance \emph{C} = 2.6 nF [Fig.~\ref{fig8}(a)], is calculated to be $f_{J}$ = 3.8 GHz. By considering the definition of current, $I_{A}\equiv dQ/dt=2e\times dn/dt \sim 2enf_{J}$, the number of the 2\emph{e}-pairs that participates in this process is determined as approximately \emph{n} $\sim$ 1. Thus, the SCRQ externally flushes out a single 2\emph{e}-pair at the rate of $f_{J}$.

In the QPS regime, the transfer of a single 2\emph{e}-pair through a JJ corresponds to a single hopping event of $\Phi_{0}$ due to the principle of duality~\cite{Moo2006}. Where, the hopping rate of $\Phi_{0}$, i.e., $f_{J}$ = 3.8 GHz, has already been utilized to normalize the QPS-induced discretized voltage [see the inset of Fig.~\ref{fig7}(a)]. By taking into account that a hopping of $\Phi_{0}$ from one thru-hole to another is phenomenologically identical to the annihilation of $\Phi_{0}$ in a certain thru-hole, $\delta$\emph{F} can be regarded as the condensation energy held in $\Phi_{0}$. In other words, we assume that the annihilation of a single $\Phi_{0}$ produces $\delta$\emph{F} that forces out a single 2\emph{e}-pair through a JJ. Thus, by assuming $\Phi_{0}$ to be the thermodynamic critical field, $\delta$\emph{F} can be expressed as $\delta F=V_{0}\times\frac{1}{2}\mu_{0}H_{\Phi_{0}}^{2}$ [Section 2.3 in Ref.~\cite{Tin2004}], where $V_{0}$ is the volume of the aerial pancake with a 150-nm thickness and 19.8-$\si{\micro}$m diameter, and $H_{\Phi_{0}}$ is the magnetic field created by a single $\Phi_{0}$ in the pancake, i.e., $\mu_{0}H_{\Phi_{0}}$ = 6.7 $\si{\micro}$T. In conclusion, $\delta F=8.3\times 10^{-22}$ J. By taking into account that the process proceeds at the rate of $f_{J}$ (3.8 GHz), the time-averaged energy produced is 3 pW. Some of the energy is consumed as work done by $E_{D}$ or is dissipated at the boundary between the SCRQ and the electrodes with a finite barrier strength [details can be referred from Fig. S6(b) in the \hyperref[supplmater]{supplementary material}]. The rest of $\delta$\emph{F} is available as driving power and is currently limited to 1 pW as shown in Fig.~\ref{fig9}(c).

It should be noted that the estimated $\delta$\emph{F} corresponds to the thermal energy of 60 K that is the temperature at which the onset of the zero-resistance state originates [see Fig.~\ref{fig2}]. Thus, an alternative explanation to that of QPS, is that the SCRQ externally releases the condensed energy that is gained by the phase transition at 60 K. However, the energy is not used up in one shot due to the sustainable p-Semi--SC junction system in which \emph{S} never increases.

\section{CONCLUSION}
\label{sec:conclusion}
A periodically perforated Nb film hosts superconductivity after repeated temperature cycling, and the superconductivity survives in the atmosphere. Zero resistance and a critical magnetic field of 12 T at 300 K are confirmed by \emph{R}--\emph{T} and \emph{V}--\emph{I} measurements, respectively. Contribution of \emph{d} electrons to superconductivity is proven by the Friedel sum rule and the $V^{2}$-dependent $dI/dV$. By considering the critical current $I_{C}$ (18.8 mA) that follows the Ambegaokar--Baratoff formula, we conclude that the nature of the RTSC is a 2D square array of \emph{d}-orbital-filled Mott-insulating JJs, which is a crystallographic analog of a CuO\textsubscript{2} plane in an HTSC. The long lattice constant of the artificial HbC results in an extraordinarily long coherence length $\xi$ of the RTSC ($\sim$20 $\si{\micro}$m), which is verified in the SC-specific phenomenon of the beating heart of PSC [Fig.~\ref{fig7}(b)]. The electrically and optically confirmed large $\mathnormal{\Delta}$ (0.8 eV) indicates that the RTSC is far from a BCS-type superconductor. By taking into account that $E_{J}$ \emph{vs.} $E_{C}$ governs this 2D superconducting system, the RTSC seems rather close to a granular superconductor which consists of inhomogeneous macroscopic grains~\cite{Sch2012,Pre2016,Sim1994,Bra1984}. In this RTSC, however, the macroscopic ``grains'' are homogeneously arrayed in an orderly square [see Fig.~\ref{fig5}], and therefore, the superconducting yield is not spatially limited.

In a 2D superconducting system, the penetration depth $\lambda_{\perp}$, which is a characteristic length for magnetic fields perpendicular to a film, is expressed as $\lambda_{\perp}=\Phi_{0}/(2\pi\mu_{0}\times I_{C})$~\cite{Tin2004,Phi1993}, where $\Phi_{0}$ is the flux quantum. By substituting 18.8 mA for $I_{C}$, $\lambda_{\perp}$ of the RTSC is calculated to be 14 nm. The relation of $\lambda_{\perp}\ll \xi$ suggests that this RTSC is a type-I superconductor. (Type-II superconductivity of the constituent Nb atoms is not manifested in the atmosphere.) In fact, a preliminary experiment using magnetic force microscopy (MFM) shows the totally blank MFM image, indicating the type-I superconductivity of this RTSC [see ``MFM Results'' in the \hyperref[supplmater]{supplementary material} for details]. In the future, Meissner diamagnetism will be addressed after eliminating the problem created by the massive silicon (Si) substrate that supports the physically microfabricated, tiny RTSC film. Although diamagnetism of a Si is 10\textsuperscript{5}-smaller than the Meissner value, the volume of the Si substrate is 10\textsuperscript{6}-larger than that of the RTSC film, veiling the Meissner effect.

Once a large current exceeding $I_{C}$ is applied, the SC phase disappears, and the bizarre, origin-asymmetry mixture of metallic and insulating states is exhibited [inset, Fig.~\ref{fig4}]. By taking into account the nature of the RTSC (i.e., as an array of \emph{d}-orbital-filled JJs), we assume that the excess current forces out some \emph{d} electrons; in other words, holes are excessively injected into the JJs, which results in the bizarre state. Although additional temperature cycles succeed in re-opening the $\mathnormal{\Delta}$ [Fig.~\ref{fig6}], the \emph{V}--\emph{I} characteristics of the emergent new condensed phase is still bizarrely asymmetric [Fig.~\ref{fig7}]; that is, the voltage increases stepwise in the positive current region, and, in contrast, it is discretized in the negative current region. Both of them are well-known SC-specific phenomena of the beating heart of PSC~\cite{Tin2004,Mey1972} and QPS~\cite{Moo2006}, respectively. (Although the co-existence of the two significant phenomena in a sole material has not been reported previously, it is what it is; we cannot do anything about it.) The DC electrical measurements confirm that therein lies a special Schottky interface consisting of the p-type semiconducting and superconducting states [Fig.~\ref{fig8}(c)]. At the interface, the charge concentration is not in equilibrium, and therefore, holes on the p-Semi side diffuse in accordance with the second law of thermodynamics, which creates the built-in electric field that accelerates an electron to the SC side. However, the accelerated electron is Andreev-reflected back as a hole to the p-Semi side with the probability unity [see the bottom-left inset of Fig.~\ref{fig8}(b): $|dI/dV|=G_{0}$]. That is, the initial nonequilibrium charge concentration is never dissolved, and the charge transport never achieves thermodynamic equilibrium. Thus, this study challenges the forbidden perpetual motion machine (PMM), and, in fact, the low-loss current flowing into an external circuit is confirmed using a very simple discharge test [Fig.~\ref{fig9}].

The refusal of a PMM is based on practical experience, but it confirms the secure, universal law of energy conservation. To create a PMM, however, we do not have to trick and/or violate the law. This universe is originated from the vacuum releasing a tremendous amount of energy suddenly~\cite{Sat1981,Gut1981}, and the beginning is frequently described as the ``ultimate free lunch.'' Still, it does not contradict the law of energy conservation. It just obeys the theory of spontaneous symmetry breaking (SSB)~\cite{Nam1960}. Thus, we already know that energy can be extracted from the vacuum at the moment of SSB. (Note that the SC-phase transition was the key for Nambu to develop the theory of SSB which led to the theory of the inflationary universe.) Of course, its beginning is not repeated persistently and therefore is not a PMM. On the other hand, the p-Semi--SC interface in this study satisfies the one and only condition for entropy never to increase [see Fig.~\ref{fig8}(c): the reversible, isolated system]. (Note that Fermi never says, ``Entropy does increase.'' He says, ``it never decreases,'' in his textbook~\cite{Fer1937}.) And, thanks to SSB, the charged particles involved in the sustainable transport process gain condensation energy each time they enter the SC phase, as discussed in the \hyperref[sec:iiic]{Section III. C}. Thus, the principle of sustainable SSB, which is realized in the p-Semi--SC junction system, may release forbidden PMMs. The SC-requiem, which is produced by making a hole in the \emph{Holy Grail}~\cite{And2011}, serendipitously generates a low-loss current spontaneously, as if the \emph{æternam contents} continue leaking out of the hole. The currently tiny amount of sub-nA can be improved soon because investigation into the ``ultimate free wine'' is undoubtedly exciting, and the improvement can be a game changer against rapid global warming and/or an alternative solution to the potential disaster, nuclear power.
\end{spacing}

\section*{Supplementary Material}\label{supplmater}
\vspace{-0.4cm}
\noindent
The supplementary material is included in this arXiv submission as an Ancillary file. The supplementary material includes the following: Materials and Methods, Figs. S1 to S11, Tables S1 to S4, MFM Results, and Full Reference List.
\vspace{-0.4cm}

\section*{Data Availability}
\vspace{-0.4cm}
\noindent
The raw data of all experimental results in this study is available in the open-access repository Dryad~\cite{Zen2020}. Also, the GDSII file of the i-line lithography used for forming the PnC pattern on a Nb film has been deposited therein. (They are also included in this arXiv submission as Ancillary files.) The author assumes no responsibility for any problem that may result from dealing with this data. In particular, the addition of the author as a responsible author and/or an inventor in any publication, including electronic publications, is prohibited.
\vspace{-0.4cm}

\section*{Acknowledgments}
\vspace{-0.4cm}
\noindent
The author thanks M. Yamawaki, Y. Higashi, K. Makise, J. Tominaga, and T. Nakano for stimulating discussions, I. Shibata and S. Yeh at the Micro-Nano Open Innovation Center (MNOIC) for the HF dry etching operation, and M. Yamazaki at the Nano-Processing Facility (NPF) for the MFM operation.

\end{document}